\documentclass[aip,apl,preprint]{revtex4-1}
\usepackage{graphicx}
\usepackage{natbib}
\usepackage{amsmath}
\usepackage{amssymb}
\usepackage{bm}
\usepackage{color}

\begin{document}

\title{Ultrafast carrier and phonon dynamics in Bi$_2$Se$_3$ crystals}

\author{J. Qi}
\email{jingbo@lanl.gov}
\affiliation{School of Physics, Georgia Institute of Technology, Atlanta, GA 30332, USA}
\affiliation{National High Magnetic Field Laboratory, Tallahassee, FL 32310, USA}

\author{X. Chen}
\affiliation{School of Physics, Georgia Institute of Technology, Atlanta, GA 30332, USA}

\author{W. Yu}
\affiliation{School of Physics, Georgia Institute of Technology, Atlanta, GA 30332, USA}

\author{P. Cadden-Zimansky}
\affiliation{National High Magnetic Field Laboratory, Tallahassee, FL 32310, USA}

\author{D. Smirnov}
\affiliation{National High Magnetic Field Laboratory, Tallahassee, FL 32310, USA}

\author{N. H. Tolk}
\affiliation{Department of Physics, Vanderbilt University, Nashville, TN 37235, USA}

\author{I. Miotkowski}
\affiliation{Department of Physics, Purdue University, West Lafayette, IN 47907, USA}

\author{H. Cao}
\affiliation{Department of Physics, Purdue University, West Lafayette, IN 47907, USA}

\author{Y. P. Chen}
\affiliation{Department of Physics, Purdue University, West Lafayette, IN 47907, USA}

\author{Y. Wu}
\affiliation{Department of Physics, Fudan University, Shanghai 200433, China}

\author{S. Qiao}
\affiliation{Department of Physics, Fudan University, Shanghai 200433, China}

\author{Z. Jiang}
\email{zhigang.jiang@physics.gatech.edu}
\affiliation{School of Physics, Georgia Institute of Technology, Atlanta, GA 30332, USA}


\begin{abstract}
Ultrafast time-resolved differential reflectivity of Bi$_2$Se$_3$ crystals is studied using optical pump-probe spectroscopy. Three distinct relaxation processes are found to contribute to the initial transient reflectivity changes. The deduced relaxation timescale and the sign of the reflectivity change suggest that electron-phonon interactions and defect-induced charge trapping are the underlying mechanisms for the three processes. After the crystal is exposed to air, the relative strength of these processes is altered and becomes strongly dependent on the excitation photon energy.
\end{abstract}

\maketitle

Recently there has been an emerging interest in Bi-based compounds, such as Bi$_{1-x}$Sb$_x$, Bi$_2$Se$_3$, and Bi$_2$Te$_3$, as exemplars of a new class of quantum matter known as a topological insulator \cite{zhang_TIreview_2010}. Among these compound materials, Bi$_2$Se$_3$ stands out owing to its relatively simple topological surface states and its sizable bulk band gap of $\sim$0.3 eV \cite{hasan_BiSe_2009,zhang_BiSe_2009}, which hold promise for applications in future spintronic and quantum computational devices. Indeed, in the past year the electronic properties of Bi$_2$Se$_3$ have attracted considerable attention \cite{cava_trans_2009,basov_IR_2010,paglione_trans_2010,fisher_trans_2010} as field effect transistor type devices have been demonstrated experimentally \cite{herrero_FET_2010,ong_FET_2010,lu_FET_2010}. However, in order to implement room-temperature operation of Bi$_2$Se$_3$-based devices, it is essential to understand the dynamics of phonons and excited carriers in this material \cite{shah_ultrafast_1999}, particularly the electron-electron, electron-phonon and phonon-phonon interactions. In this Letter, we report the ultrafast time-resolved optical spectroscopy study of Bi$_2$Se$_3$ crystals in both the time domain and the energy domain. Our measurements reveal three underlying relaxation processes in the transient response of Bi$_2$Se$_3$, each associated with different physical mechanisms. It is also shown that the relative strength of these processes is sensitive to air exposure of the samples. The observed charge trapping and air doping effects are likely due to the presence of Se vacancies, a major issue material scientists working to use the novel properties of Bi$_2$Se$_3$ will face in the near term.

The Bi$_2$Se$_3$ single crystals studied in this work were synthesized via the Bridgman method at Purdue University and Fudan University. During crystal growth, the mixture of high purity elements was first deoxidized and purified by multiple vacuum distillations, and then heated to 850-900 $^\circ$C for 15 hours, followed by a slow cool down under a controlled pressure of Se to compensate for possible Se vacancies. Afterwards, the samples were zone refined at a speed of 0.5-1.5 mm/hour with a linear temperature gradient set to 4-5 $^\circ$C/cm, until a temperature of 670 $^\circ$C was reached. The as-grown Bi$_2$Se$_3$ crystals from both groups are naturally n-doped due to remnant Se vacancies \cite{cava_trans_2009}. Hall measurements show typical carrier densities over $1\times 10^{18}\ $cm$^{-3}$ at ambient conditions.

In the pump-probe experiment, the transient reflectivity changes, $\Delta R/R$, are measured at room temperature by employing a Ti:Sapphire laser with a repetition rate of 76 MHz. The laser is capable of producing $\sim$150 fs-wide pulses in a wavelength range from $\sim$720 nm (1.72 eV) to $\sim$940 nm (1.32 eV). Cross-polarized (linear) pump and probe beams of the same wavelength are focused onto the sample in a spot $\sim$100 $\mu$m in diameter. To avoid heating damage, the typical pump light fluence used in this work is 2.5 $\mu$J/cm$^2$.
\begin{figure}
\includegraphics[width=8cm]{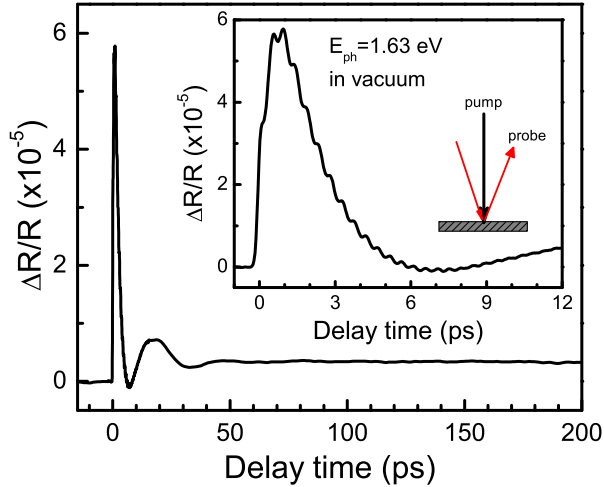}
\caption{\label{fig:deltaR} Time-resolved response of Bi$_2$Se$_3$ at room temperature for the pump-probe photon energy 1.63 eV over long (main panel) and short (inset) timescales. A schematic drawing of the measurement configuration is also shown in the inset.}
\end{figure}

\begin{figure}
\includegraphics[width=8cm]{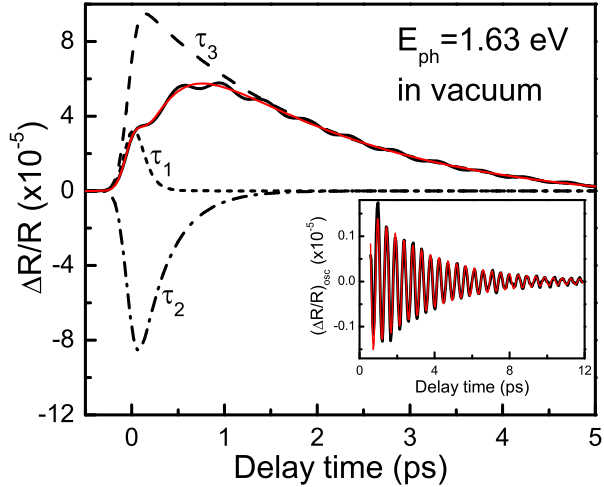}
\caption{\label{fig:oscillation}(color online) $\Delta R/R$ signal of Bi$_2$Se$_3$ as a function of delay time at $E_{ph}=1.63$ eV. The best fit to data (red solid line) is obtained by using a model consisting of three independent exponential relaxation processes with time constant $\tau_1$ (short dashed line), $\tau_2$ (dashed-dot line), and $\tau_3$ (dashed line). Inset: Extracted oscillations from the measured $\Delta R/R$ and the best fit to data (red solid line) as an exponentially damped oscillation.}
\end{figure}

Figure \ref{fig:deltaR} shows the measured $\Delta R/R$ with the pump-probe wavelength centered at 760 nm (1.63 eV) over long (main panel) and short (inset) timescales. As one can see, $\Delta R/R$ undergoes a sharp rise within $\sim$1 ps, followed by a decay. In addition, two apparent oscillatory behaviors are superimposed on the $\Delta R/R$ profile: One appears within the initial $\sim$15 ps with a higher frequency as evidenced in the small periodic ``wiggles'' in the inset to Fig. \ref{fig:deltaR}, while the other arrives after $\sim$6 ps and decays in a few periods with a much lower frequency but a larger amplitude (see the main panel of Fig. \ref{fig:deltaR}). The low frequency oscillation has been widely observed in ultrafast time-resolved optical measurements of other systems \cite{qi_mechanical_2010} and has been attributed to coherent longitudinal acoustic phonons generated by the pump light. To study the high frequency oscillation, we first extract it by subtracting the non-oscillatory component from the differential reflectivity data, and then fit it as an exponentially damped oscillation \cite{xu_fitting_2008}. As shown in the inset to Fig. \ref{fig:oscillation}, our data can be fit well by this method with a single frequency $\nu$ centered at 2.13 THz and a damping rate of $\Gamma\sim0.33$ THz. Here the deduced $\nu$ value is consistent with the energy of the $A^{1}_{1g}$ longitudinal optical phonon in Bi$_2$Se$_3$ obtained by Raman spectroscopy \cite{becher_raman_1977}, while the damping rate represents the timescale of the phonon-phonon interactions. Moreover, $\nu$ and $\Gamma$ are found to be independent of the excitation photon energy $E_{ph}$ within our experimental error.

\begin{figure}
\includegraphics[width=8cm]{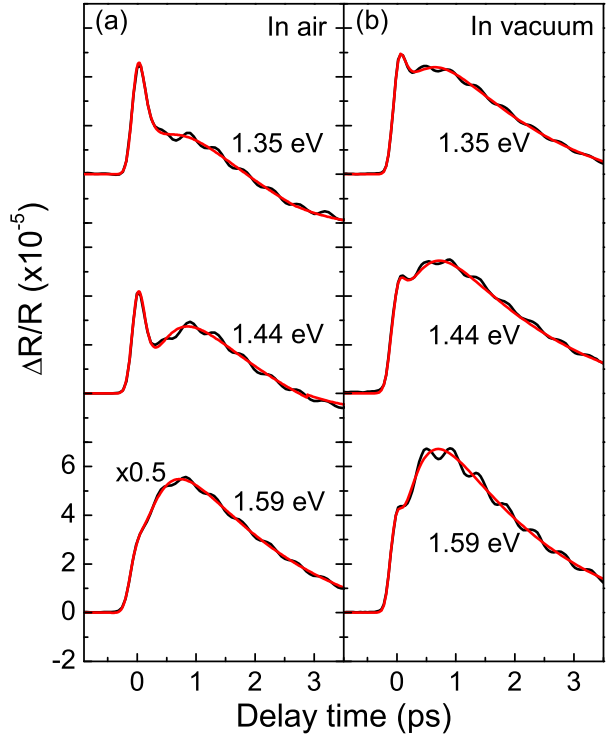}
\caption{\label{fig:deltaR_air_vac}(color online) Side-by-side comparison of the $\Delta R/R$ signals of Bi$_2$Se$_3$ exposed to air and in vacuum at three different excitation photon energies. Traces are offset for clarity. The red solid lines are the best fits to data with three independent exponential relaxation processes.}
\end{figure}

Now let us discuss the non-oscillatory transient response of Bi$_2$Se$_3$ in the first few picoseconds after pumping. Figure \ref{fig:oscillation} plots $\Delta R/R$ as a function of delay time measured at room temperature and $E_{ph}=1.63$ eV. We note that the reflectance rise time in Bi$_2$Se$_3$ ($\sim$1 ps) is much longer than that expected for typical semiconductors and metals ($\sim$300 fs)\cite{perakis_htc_2001}.  This timescale corresponds to the delayed thermalization of the carriers via electron-electron interactions. Such a slow rise may indicate the presence of a relaxation process with a negative amplitude ($A_2$), as illustrated by the dashed-dot line in Fig. \ref{fig:oscillation}. In addition, as further evidenced in Fig. \ref{fig:deltaR_air_vac}, two positive processes also contribute to the transient reflectivity changes, with distinctive time constant $\tau_1$ and $\tau_3$. Thus, to obtain the best fit to the data in Figs. \ref{fig:oscillation} and \ref{fig:deltaR_air_vac} (the red solid lines), we employ a model consisting of three independent exponential relaxation processes \cite{Xu_MgB2_2003}. The deduced time constants, $\tau_1=0.11\pm0.02$ ps, $\tau_2=0.32\pm0.03$ ps, and $\tau_3=2.3\pm0.2$ ps, are independent of the pump intensity and the excitation photon energy available with our experimental set-up.

What is the possible origin of the three relaxation processes? The first process is characterized by a fast relaxation ($\tau_1=110$ fs), as the excited carriers lose their high excess energy rapidly through the electron-phonon (particularly electron-optical-phonon) interactions \cite{Xu_MgB2_2003,allen_PRL_1987,taneda_FS_2007}. The third process, on the other hand, represents a relatively slow relaxation ($\tau_3=2.3$ ps), which might be attributed to electron-acoustic-phonon interactions \cite{Xu_MgB2_2003,taneda_FS_2007}. The co-existence of a sub-ps and a few-ps electron-phonon relaxation processes due to different phonon modes has been reported previously, and quantitative agreement is achieved in the extracted relaxation timescales \cite{Xu_MgB2_2003,taneda_FS_2007}. The second process with a negative amplitude $A_2$, however, is more subtle. Such a negative-amplitude process was previously observed in strongly correlated electron systems \cite{taylor_negative_2006}, as well as in disordered low-temperature molecular-beam-epitaxy grown III-V semiconductors \cite{calawa_LTGaAs_1991,wang_IIIMnV_2006}. One possibility is that this process is associated with the ultrafast trapping of electrons by Se vacancies. This trapping mechanism could cause a negative change in reflectivity and limit the relaxation time to less than 1 ps \cite{wang_IIIMnV_2006}, in agreement with our observations.
\begin{figure}
\includegraphics[width=8cm]{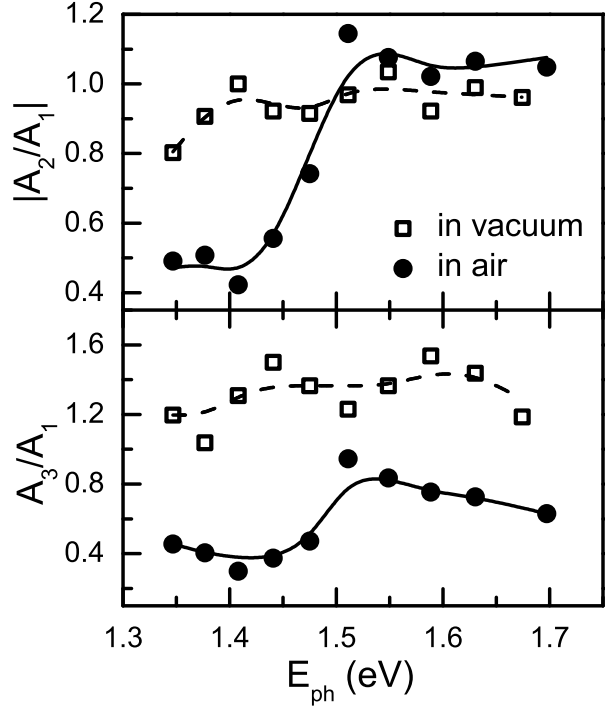}
\caption{\label{fig:amplitude_energy} $|A_2/A_1|$ and $A_3/A_1$ as a function of $E_{ph}$ for Bi$_2$Se$_3$ crystals exposed to air (filled dots) and in vacuum (open squares). Solid and dashed lines are the guides to the eye.}
\end{figure}

An important material issue for Bi$_2$Se$_3$ crystals is unintentional charge doping from air, which could modify or even dominate its electronic properties \cite{fisher_trans_2010}. To pursue this issue, we perform comparative studies of the time-resolved response of Bi$_2$Se$_3$ exposed to air and in vacuum at different excitation photon energies. To prepare an ``in-vacuum'' sample, we first rapidly cleave it to obtain a mirror-like surface and load it into a sealed measurement chamber within 20 seconds, the chamber is then pump down to $\sim$$10^{-5}$ mbar in $\sim$20 mins. As shown in Fig. \ref{fig:deltaR_air_vac}, a clear difference in $\Delta R/R$ is evident as a function of $E_{ph}$. Since $\tau_{1,2,3}$ exhibit almost no change when the samples are exposed to air, the observed differences are mainly from the relative changes in the amplitudes $A_{1,2,3}$. Figure \ref{fig:amplitude_energy} summarizes the $E_{ph}$ dependence of $|A_2/A_1|$ and $A_3/A_1$ for the measurements in air and in vacuum. While the relative strength of the three relaxation processes in vacuum is only weakly dependent on $E_{ph}$, an abrupt change occurs for the samples exposed to air when $E_{ph}\sim1.47$ eV. Air doping may raise the Fermi level in Bi$_2$Se$_3$ and affect the band bending near the crystal surface. This alters the spatial electron-hole distribution function \cite{cardona_book_2000}, which impacts the ultrafast dynamics of the material. Further experimental and theoretical work is needed to understand our observation. Finally, we find that the extracted $A^{1}_{1g}$ optical phonon energy ($\nu=2.13$ THz) from the in-vacuum samples is slightly, but consistently, lower than that obtained for the samples exposed to air (2.15 THz) at all $E_{ph}$ measured. Given that the theoretical value of this energy is 2.03 THz for a perfect Bi$_2$Se$_3$ crystal\cite{becher_raman_1977}, Se vacancies may also underlie the observed difference.

In conclusion, we have performed ultrafast time-resolved optical spectroscopy measurements on Bi$_2$Se$_3$ crystals and observed the contributions from three exponential relaxation processes with distinctive time constants. The origin of the three processes can be attributed to electron-phonon interactions and defect-induced charge trapping. Our work also shows that air exposure can significantly effect the carrier and phonon dynamics in Bi$_2$Se$_3$ crystals by altering the relative strength of these three processes.

We would like to thank W. Y. Ruan and M.-Y. Chou for helpful discussions. This work is supported by the DOE (DE-FG02-07ER46451). A portion of this work was performed at the National High Magnetic Field Laboratory, which is supported by NSF Cooperative Agreement No. DMR-0654118, by the State of Florida, and by the DOE.


\begin{thebibliography}{text}
\bibitem{zhang_TIreview_2010}For a recent review, see X.-L. Qi and S.-C. Zhang, arXiv:1008.2026.
\bibitem{hasan_BiSe_2009}Y. Xia, D. Qian, D. Hsieh, L. Wray, A. Pal, H. Lin, A. Bansil, D. Grauer, Y. S. Hor, R. J. Cava, and M. Z. Hasan, Nature Physics {\bf 5}, 398 (2009).
\bibitem{zhang_BiSe_2009}H. Zhang, C.-X. Liu, X.-L. Qi, X. Dai, Z. Fang, and S.-C. Zhang, Nature Physics {\bf 5}, 438 (2009).
\bibitem{cava_trans_2009}Y. S. Hor, A. Richardella, P. Roushan, Y. Xia, J. G. Checkelsky, A. Yazdani, M. Z. Hasan, N. P. Ong, and R. J. Cava, Phys. Rev. B {\bf 79}, 195208 (2009).
\bibitem{basov_IR_2010}A. D. LaForge, A. Frenzel, B. C. Pursley, T. Lin, X. Liu, J. Shi, and D. N. Basov, Phys. Rev. B {\bf 81}, 125120 (2010).
\bibitem{paglione_trans_2010}N. P. Butch, K. Kirshenbaum, P. Syers, A. B. Sushkov, G. S. Jenkins, H. D. Drew, and J. Paglione, Phys. Rev. B {\bf 81}, 241301(R) (2010).
\bibitem{fisher_trans_2010}J. G. Analytis, R. D. McDonald, S. C. Riggs, J.-H. Chu, G. S. Boebinger, and I. R. Fisher, arXiv:1003.1713.
\bibitem{lu_FET_2010}J. Chen, H. J. Qin, F. Yang, J. Liu, T. Guan, F. M. Qu, G. H. Zhang, J. R. Shi, X. C. Xie, C. L. Yang, K. H. Wu, Y. Q. Li, and L. Lu, arXiv:1003.1534.
\bibitem{herrero_FET_2010}H. Steinberg, D. R. Gardner, Y. S. Lee, and P. Jarillo-Herrero, arXiv:1003.3137.
\bibitem{ong_FET_2010}J. G. Checkelsky, Y. S. Hor, R. J. Cava, and N. P. Ong, arXiv:1003.3883.
\bibitem{shah_ultrafast_1999}J. Shah, \textit{Ultrafast spectroscopy of semiconductors and semiconductor nanostructures} (Springer, 1999).
\bibitem{qi_mechanical_2010}J. Qi, J. A. Yan, H. Park, A. Steigerwald, Y. Xu, S. N. Gilbert, X. Liu, J. K. Furdyna, S. T. Pantelides, and N. Tolk, Phys. Rev. B {\bf 81}, 115208 (2010); and the references therein.
\bibitem{xu_fitting_2008}A. Q. Wu, X. Xu, and R. Venkatasubramanian, Appl. Phys. Lett. {\bf 92}, 011108 (2008).
\bibitem{becher_raman_1977}W. Richter, H. K\"{o}hler, and C. R. Becker, Physica Status Solidi B-Basic Research {\bf 84}, 619 (1977).
\bibitem{perakis_htc_2001}S. Rast, M. L. Schneider, M. Onellion, X. H. Zeng, W. Si, X. X. Xi, M. Abrecht, D. Ariosa, D. Pavuna, Y. H. Ren, G. L\"{u}pke, and I. Perakis, Phys. Rev. B {\bf 64}, 214505 (2001).
\bibitem{Xu_MgB2_2003} Y. Xu, M. Khafizov, L. Satrapinsky, P. Kus, A. Plecenik, and R. Sobolewski, Phys. Rev. Lett. {\bf 91}, 197004 (2003).
\bibitem{allen_PRL_1987}P. B. Allen, Phys. Rev. Lett. {\bf 59}, 1460 (1987).
\bibitem{taneda_FS_2007}T. Taneda, G. P. Pepe, L. Parlato, A. A. Golubov, and R. Sobolewski, Phys. Rev. B {\bf 75}, 174507 (2007).
\bibitem{taylor_negative_2006}J. Demsar, V. K. Thorsm{\o}lle, J. L. Sarrao, and A. J. Taylor, Phys. Rev. Lett. {\bf 96}, 037401 (2006).
\bibitem{calawa_LTGaAs_1991}S. Gupta, M. Y. Frankel, J. A. Valdmanis, J. F. Whitaker, G. A. Mourou, F. W. Smith, and A. R. Calawa, Appl. Phys. Lett. {\bf 59}, 3276 (1991).
\bibitem{wang_IIIMnV_2006}J. G. Wang, C. J. Sun, Y. Hashimoto, J. Kono, G. A. Khodaparast, \L. Cywi\'{n}ski, L. J. Sham, G. D. Sanders, C. J. Stanton, and H. Munekata, J. Phys: Condens. Matter {\bf 18} R501 (2006).
\bibitem{cardona_book_2000}T. Dekorsy, G. C. Cho, and H. Kurz, Chapter 4 in \textit{Light Scattering in Solids VIII: Fullerenes, Semiconductor Surfaces, Coherent Phonons (Topics in Applied Physics, Volume 76)} (Springer, 2000).
\end{thebibliography}
\end{document}